# Silicon-On-Insulator Lateral Dual Sidewall Schottky (SOI-LDSS) Concept for Improved Rectifier Performance: A Two-Dimensional Simulation Study

**M. Jagadesh Kumar[1] and C. Linga Reddy**

Department of Electrical Engineering

Indian Institute of Technology, Delhi

Hauz Khas, New Delhi – 110 016, INDIA

Email: mamidala@ieee.org    FAX: 91-11-2658 1264

***Abstract:*** In this paper, we demonstrate that the performance of silicon Schottky rectifiers on SOI can be significantly improved using a Lateral Dual Sidewall Schottky (LDSS) concept. Our results based on numerical simulation show that the LDSS structure on SOI has low forward voltage drop and low reverse leakage current while its breakdown voltage is significantly larger than that of a conventional Schottky rectifier.

Key Words: Schottky rectifier, forward drop, leakage current, SOI, simulation

---

[1] Author for correspondence

# 1.INTRODUCTION

Power switching devices constitute the heart of modern power electronic systems. While there is a large choice of power switching devices, Schottky rectifiers are known for their absence of reverse recovery problems[1-5]. They are very commonly used in switched mode power supplies as the efficiency of these power supplies can be significantly improved since the forward voltage drop of the Schottky rectifier is very small in relation to the output voltage amplitude. In the recent past, there have been a number of novel Schottky rectifiers reported in literature [6-10]. For example, the concept of using two metals with different workfunctions  in a Schottky rectifier[6-9] or using a sidewall Schottky contact [4] is very interesting owing to their excellent suitability for high frequency low power applications. In these structures, the low barrier Schottky contact conducts during the forward bias and is pinched off during the reverse bias by a high-barrier metal so that the reverse leakage current corresponds to that of the high barrier Schottky contact.    In this paper, the Lateral Dual Sidewall Schottky (LDSS) concept is applied  to a silicon based Schottky rectifier on SOI and evaluate its performance against a silicon Lateral Conventional Schottky (LCS) rectifier is evaluated.  Based on two-dimensional simulation results, we demonstrate that the forward characteristics of the proposed LDSS rectifier on SOI are far superior to that of the lateral conventional Schottky rectifier. The detailed analysis and the reasons for the improved performance of the proposed LDSS rectifier on SOI are discussed  in the following sections.

## 2. DEVICE STRUCTURE AND PARAMETERS

Cross sectional views of the proposed Lateral Dual Sidewall Schottky (LDSS) rectifier on SOI and the lateral conventional Schottky (LCS) structure on SOI implemented using two dimensional device simulator MEDICI [11] are shown in Fig. 1. Details on the fabrication procedure to obtain the dual side-wall Schottky contact can be found in [9]. In the case of the LDSS structure, the anode consists of both the low-barrier Schottky contact as well as high-barrier Schottky contact. Nickel (Schottky barrier height $\phi_{BL} = 0.57$ eV) is chosen for the low-barrier Schottky contact and Tungsten (Schottky barrier height $\phi_{BH} = 0.67$ eV) is chosen for high-barrier Schottky contact as these two are the well studied and most commonly used metals for Schottky contacts. The cathode contact is taken from the N$^+$ region. Field plate with a length of 3.5 μm is used to avoid the electric field crowding at the Schottky contact. Drift region doping (N$_D$) is chosen to be $5 \times 10^{16}$ cm$^{-3}$ and its thickness is 0.5 μm. Field oxide thickness is 0.2 μm and buried oxide thickness is 2 μm. High-barrier Schottky trench depth is chosen to be 0.4 μm and low barrier Schottky trench depth is chosen to be 0.1 μm as this combination gives the low forward voltage drop as well as less reverse leakage current. In the case of the LCS structure, the anode is made of either the high-barrier Schottky (HBS) contact (Tungsten) or the low-barrier Schottky (LBS) contact (Nickel). All the other parameters of the LCS structure are kept similar to that of the LDSS structure unless mentioned.

# 3. SIMULATION RESULTS AND DISCUSSION

Fig. 2(a) shows a comparison of the simulated forward IV characteristics of the proposed Lateral Dual Sidewall Schottky (LDSS) rectifier on SOI with the compatible Lateral Conventional Schottky (LCS) rectifier on SOI. It can be observed from the figure that the forward characteristic of the proposed LDSS rectifier is close to that of low-barrier LCS as low-barrier Schottky contact plays an essential role in providing most of its current.

Simulated reverse IV characteristics of the proposed LDSS rectifier on SOI and its compatible Lateral Conventional Schottky (LCS) rectifier on SOI are shown in Fig. 2(b). Here, we observe that the low-barrier and the high-barrier LCS rectifiers exhibit a lower breakdown voltage due to the large peak electric field at the Schottky contact as shown in Fig. 3. However, the LDSS rectifier shows not only a sharp breakdown but also a significantly improved breakdown voltage. This improvement is due to the shifting of the peak electric field from the Schottky contact to the field plate edge as shown in Fig. 3. The reverse leakage current of the proposed LDSS rectifier on SOI is close to that of high-barrier LCS rectifier as the low-barrier Schottky metal of the proposed LDSS rectifier is pinched off and high-barrier Schottky alone is responsible for the current conduction during the reverse bias.

# 4. CONCLUSION

From our simulation analysis, we conclude that the proposed Lateral Dual Sidewall Schottky (LDSS) rectifier on SOI behaves as a low-barrier Lateral Conventional Schottky (LCS) rectifier on SOI in terms of its low forward voltage drop and it behaves as a high-barrier Lateral Conventional Schottky (LCS) rectifier on SOI under reverse bias conditions. An important result is that the LDSS structure on SOI exhibits a significantly enhanced sharp breakdown compared to the LCS structure. The combined low forward voltage drop, less reverse leakage current and excellent reverse blocking capability make the proposed LDSS rectifier on SOI attractive for use in high-speed and low-loss power IC applications.

**Figure Captions**

Fig. 1   Cross-sectional view of  (a) the    lateral dual sidewall Schottky (LDSS) structure on SOI and (b) the lateral conventional Schottky (LCS) structure on SOI.

Fig. 2 (a) Forward conduction and (b) Reverse blocking characteristics of the low-barrier and high-barrier LCS and the  LDSS rectifiers.

Fig. 3 Electric field variation along the horizontal line at the field-oxide / silicon interface of low-barrier and high-barrier LCS and  the LDSS rectifiers near the breakdown voltage of low-barrier and high-barrier LCS rectifiers.

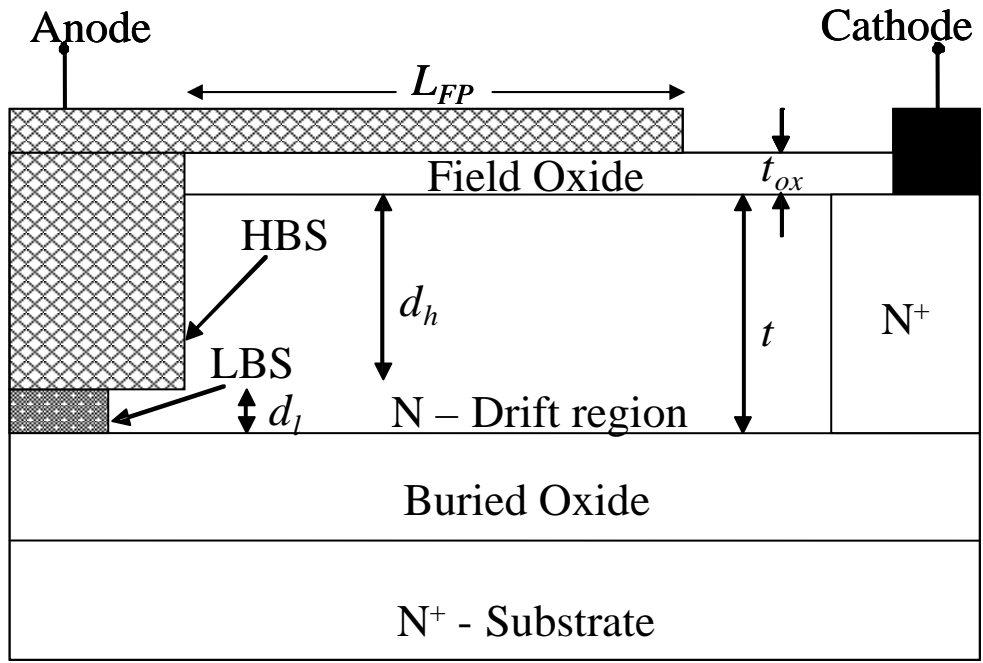

Fig. 1(a)

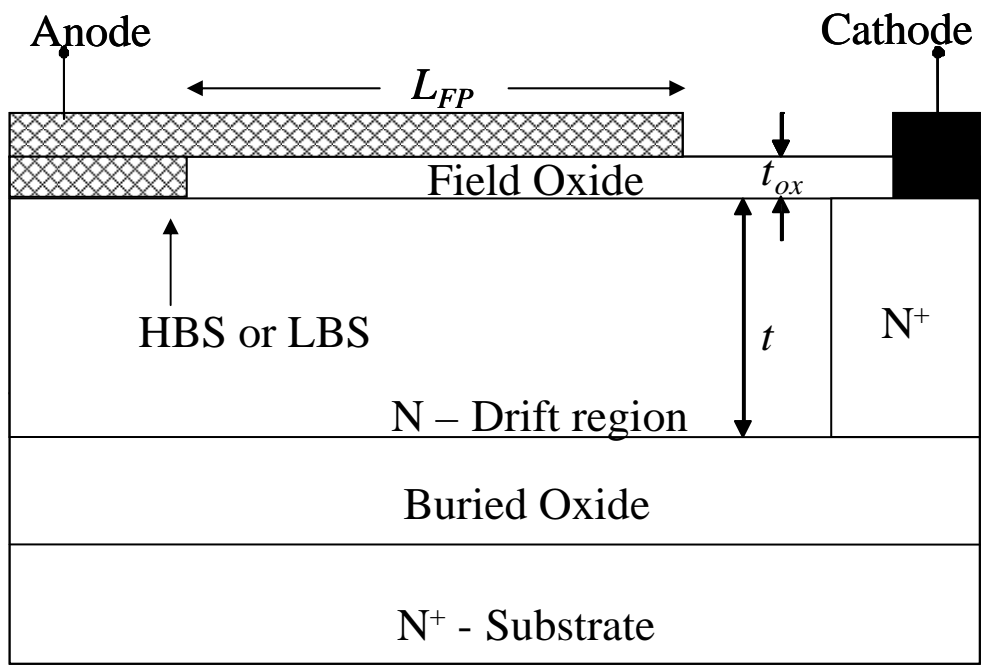

Fig. 1(b)

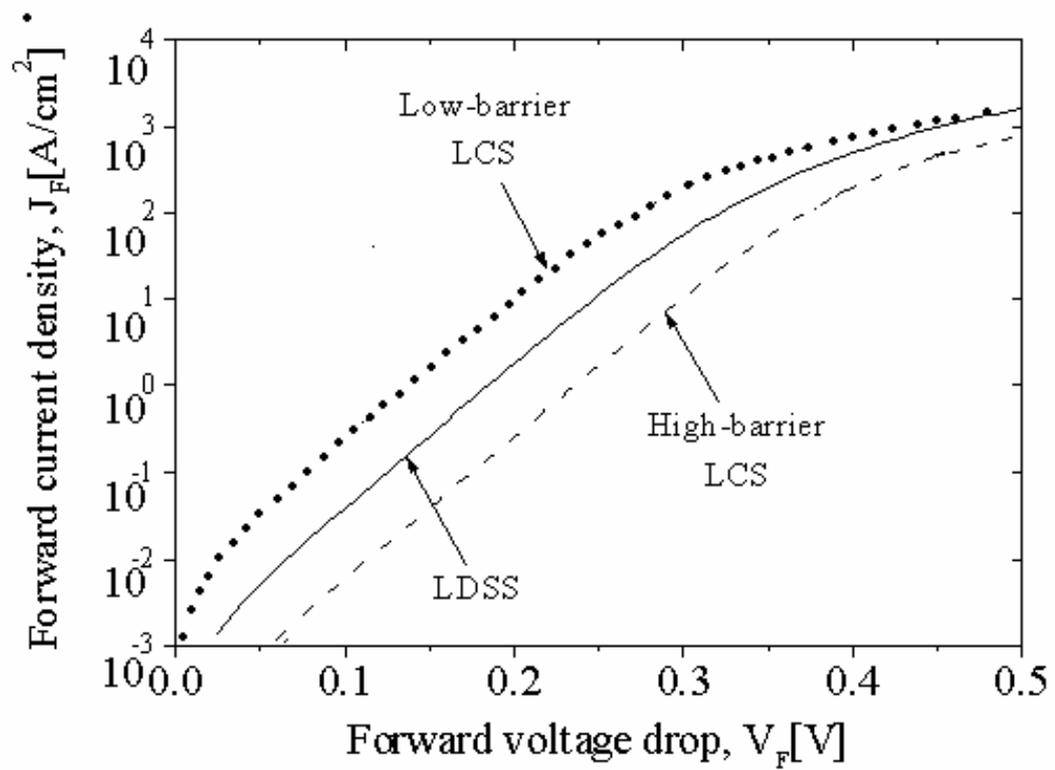

Fig. 2(a)

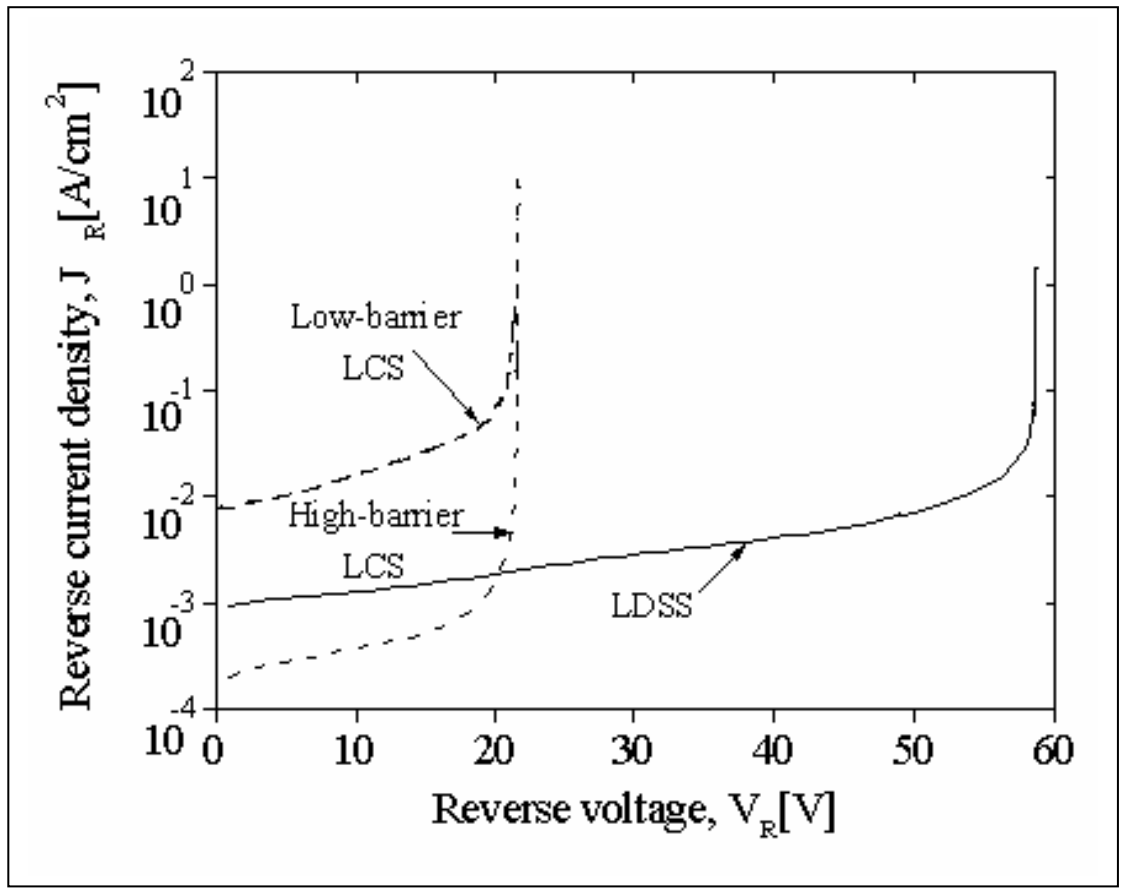

Fig. 2(b)

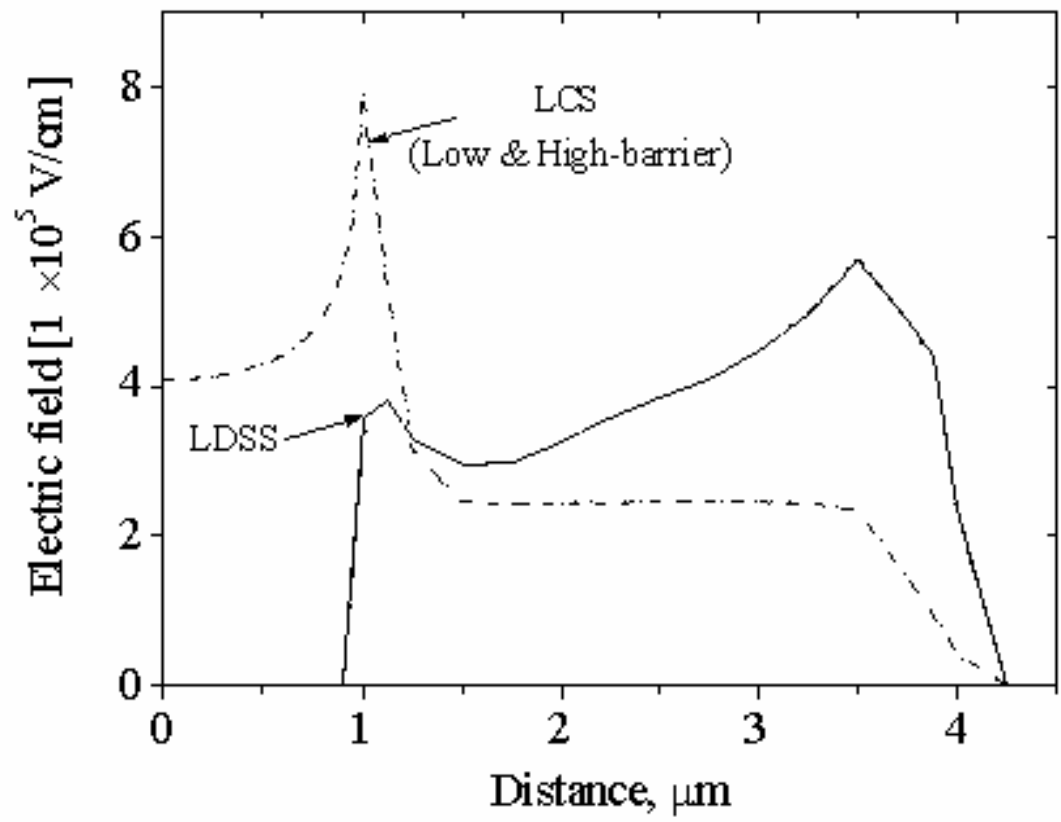

Fig. 3